\documentclass[preprint,amsmath,amssymb,superscriptaddress,showpacs,aps12pt]{revtex4}

\usepackage[utf8]{inputenc}
\usepackage{graphicx}
\usepackage{epsfig}
\usepackage{amssymb}
\usepackage{dcolumn}
\usepackage{bm}
\usepackage{hyperref}

\newcommand{\ba}{\begin{array}}
\newcommand{\ea}{\end{array}}
\def\br{\begin{eqnarray}}
\def\er{\end{eqnarray}}
\def\be{\begin{equation}}
\def\ee{\end{equation}}

\def\({\left(}
\def\){\right)}

\def\<{\left\langle}
\def\>{\right\rangle}

\begin{document}

\title{Anomalous mass dimensions and Schwinger-Dyson equations boundary condition}

\author{A. Doff}
\email{agomes@utfpr.edu.br}

\affiliation{Universidade Tecnol\'ogica Federal do Paran\'a - UTFPR - DAFIS
Av Monteiro Lobato Km 04, 84016-210, Ponta Grossa, PR, Brazil}

\affiliation{Universidade Estadual Paulista (UNESP), Instituto de F{\'i}sica Te\'orica, Rua Dr. Bento T. Ferraz, 271, Bloco II, 01140-070, S\~ao Paulo, SP, Brazil}

\author{A. A. Natale} 
\email{natale@ift.unesp.br}

\affiliation{Universidade Estadual Paulista (UNESP), Instituto de F{\'i}sica Te\'orica, Rua Dr. Bento T. Ferraz, 271, Bloco II, 01140-070, S\~ao Paulo, SP, Brazil}

\begin{abstract}
Theories with large mass anomalous dimensions ($\gamma_m$) have been extensively studied because of their deep consequences for models where the scalar bosons are composite. Large $\gamma_m$ values may appear when a non-Abelian gauge theory has a large number of fermions or is affected by four-fermion interactions. In this note we provide a simple explanation how $\gamma_m$ can be directly read out from the IR and UV boundary conditions derived from the gap equation, and verify that moderate $\gamma_m$ values appear when the theory possess a large number of fermions, but large $\gamma_m$ values are obtained only when four-fermion interactions are added to the theory. We also verify how the critical line separating the different chiral phases emerge from these conditions.  
\end{abstract}



\maketitle



\par The $125$ GeV resonance discovered at the LHC \cite{atlas} has many of the characteristics expected for the Standard Model (SM) Higgs boson. If this particle is a composite or an elementary scalar boson is still an open question. Many models have considered the possibility of a light composite Higgs based on effective Higgs potentials as reviewed in Ref.\cite{h1}. The possibility that the Higgs boson is a composite state instead of an elementary one is more akin to the phenomenon of spontaneous symmetry breaking that originated from the Ginzburg-Landau Lagrangian, which can be derived from the microscopic BCS theory of superconductivity describing the electron-hole interaction. 

\par  The possibility of spontaneous symmetry breaking promoted by a composite scalar boson formed by new fermions has been discussed with the use of many models, the technicolor (TC) being the most popular one \cite{tc1}.  However the phenomenology of these models depend crucially on these new fermions (or technifermions) self-energy. In the early models this self-energy was considered to be given by the standard operator product expansion (OPE) result \cite{politzer}:
$$
\Sigma_{TC} (p^2) \propto \frac{\left\langle {\bar{T}_f}T_f\right\rangle}{p^2} \,\, ,
$$
where $\left\langle {\bar{T}_f}T_f\right\rangle$ is the TC condensate of order of a few hundred GeV, i.e. the order of the SM vacuum expectation
value (vev). Unfortunately this behavior does lead to models with incompatibilities with the experimental data.
 A possible way out of this dilemma was proposed by Holdom \cite{holdom}, remembering that the self-energy behaves as
\be 
\Sigma_{TC} (p^2)\approx \frac{\left\langle {\bar{T}_f}T_f\right\rangle_\mu}{p^2} \left(\frac{p^2}{\mu^2}\right)^{\gamma_m /2} \,\, 
\label{eqa}
\ee
where $\mu$ is the characteristic TC scale and $\gamma_m$ the mass anomalous dimension associated to the fermionic condensate. 
As can be verified from Eq.(\ref{eqa}) a large anomalous dimension leads to a hard asymptotic self-energy and may solve the many problems of the SM symmetry breaking promoted by composite bosons. 

\par The work of Ref.\cite{holdom} started the search for theories that could present a large mass anomalous dimension, leading to fermionic
self-energies decreasing slowly with the momenta, and consequently to more realistic models of dynamical symmetry breaking (dsb). It is interesting to note that a hard asymptotic self-energy is even able to generate a scalar composite lighter 
than the scale of the SM vev \cite{us1,us2}. Models proposing such large anomalous dimensions were reviewed in Ref.\cite{yamawaki}, and theories with large anomalous dimensions ($\gamma_m$) are quite desirable for technicolor phenomenology \cite{sannino}. Studies
of these anomalous dimensions in many different non-Abelian models have been performed through analytical methods and lattice 
simulations as can be seen in Ref.\cite{g1,g2,g3,g4,g5,g6,g7} and references therein. 
The importance of these studies is not only related to the dsb model building but also to the knowledge of the different phases of non-Abelian gauge theories.

\par Early work with Schwinger-Dyson equations (SDE) in the $SU(N)$ case verified 
that $\gamma_m \approx 1$ and is not strongly affected by high order corrections \cite{g1}. Models with a slowly running
coupling, i.e. near a non-trivial fixed point, in non-Abelian gauge theories started to be studied in
Refs.\cite{apel, ap4,ap5} and seems to enhance the $\gamma_m$ values. In the Ref.\cite{apel} the fixed point was obtained from the two-loop $\beta$ function for a $SU(N)$ theory with fermions in the fundamental representation. One analysis of this problem in the case of other groups and fermionic representations can be seen in Ref.\cite{sn2}. Large mass anomalous dimensions seems also to appear naturally in what is now known as gauged Nambu-Jona-Lasinio  models, as shown in the works of Refs.\cite{yama1, yama2, mira2, yama3, mira3, yama4}. In these last type of models two coupling constants enter into action: the gauge coupling ($\alpha$) and the 4-fermion one ($g$), and there is a critical line described by a combination of these couplings where the chiral symmetry is broken. At this critical line the dynamical fermion mass behaves as a slowly decreasing function with the momentum\cite{takeuchi,kondo}, and not much different from what is expected in a theory with bare masses.

\par Limits on $\gamma_m$ can also be derived in specific models. An upper bound $\gamma_m \leq 2$ comes out from unitarity of conformal field theories \cite{g2}. Conformal bootstrap methods applied to $SU(N_f)_V$ symmetric conformal field theories suggest $\gamma_m < 1.31 $ for $N_f =8$ \cite{g3} and $\gamma_m \leq 1.29 $ for $N_f =12$ \cite{g4}.  An enormous effort has been pursued by different groups performing lattice simulations to reveal $\gamma_m$ values in $SU(3)$ with many flavors \cite{g5,g6,g7}. Some works may present
different $\gamma_m$ reflecting different approaches to determine this quantity. As one example we can quote the lattice simulation of
Ref.\cite{g7} where the anomalous dimension for $SU(3)$ with $N_f=12$ was found to be relatively small, while a SDE approach taking into account four-fermion interactions produce a larger anomalous dimension for the same model \cite{us3}. However this fact may not be a surprise, but just may indicates that
four-fermion interactions are necessarily responsible for large $\gamma_m$ values.  

\par In this note we were moved by the idea of showing in a simple way how the mass anomalous dimensions vary in different models, and
will discuss how the boundary conditions of the  anharmonic oscillator representation of the gap equation are directly related  with  the mass anomalous dimensions. We discuss how such boundary conditions (and $\gamma_m$) are  affected by inclusion of effects as a large number of fermions  (leading to what is called walking theory),  or by the inclusion of four-fermion interactions. We argue that
the anomalous dimension can be read out directly from the boundary conditions, which is a simple result although we are not
aware that this fact was stated before. We also recover the existence of the critical line, and verify how the mass anomalous
dimensions may vary with the different boundary conditions after considering the numerical solution of the fermionic  gap equation  
in the  anharmonic oscillator representation. Lastly, we verify that without four-fermion interactions the existence
of a large anomalous dimension is not compelling, whereas the opposite is true for a large range of coupling constants. 

\par The fermionic SDE in Landau gauge for a $SU(N)$ gauge theory, with fermions in the fundamental representation, can be written 
as \cite{Richard,georgi,us4}
\be
\Sigma(p)=m_0 + \frac{3C_2}{8\pi^2}\frac{\bar{g}^2(p^2)}{p^2} \int^{p}_{0}kdk \frac{k^2\Sigma(k)}{k^2+\Sigma^2(k)}
+ \frac{3C_2}{8\pi^2}\int^{\Lambda}_{p}kdk\frac{\bar{g}^2(k^2)\Sigma(k)}{k^2+\Sigma^2(k)} \, ,
\label{eq1} 
\ee 
\noindent  where $\Sigma(p)$ is the dynamical fermion mass,  $ C_2 =C_2 (F)$ is the Casimir operator for fermions in the fundamental representation and $\bar{g}^2(p^2)$ is the running coupling constant. In order to simplify the analysis, we will assume the walking limit of this equation where $\bar{g}^2(p^2) = g^2 $ is constant, in addition we also consider the set of new variables 
\br 
&&t=\ln\frac{p}{\mu} \,\,\,, \,\,\, s=\ln\frac{k}{\mu} \,\,\,,\,\,\,X(t) = \frac{\Sigma(p)}{\sqrt{p^2}} = \frac{\Sigma(t)}{\mu e^{t}}\,\,\,,\,\,\,X(s) = \frac{\Sigma(k)}{\sqrt{k^2}} = \frac{\Sigma(s)}{\mu e^{s}}.
\label{eq2}
\er 
\noindent  With these new variables, after considering the chiral limit $m_0 =0$,  Eq.(\ref{eq1}) takes the following form
\be
X(t)=  \frac{a}{2}e^{-3t}\int^{t}_{t_0}dse^{3s}\frac{X(s)}{1+X^2(s)}
+ \frac{a}{2}e^{-t}\int^{t_{\Lambda}}_{t}dse^{s}\frac{X(s)}{1+X^2(s)} \, ,
\label{eq3} 
\ee where $a=\frac{3C_2g^2}{4\pi^2}$, $t_0 = \ln\frac{p}{\mu}(p\to 0)$  and  $t_{\Lambda} = \ln\frac{\Lambda}{\mu}(\Lambda\to \infty)$. It is then possible to transform this integral equation into a differential one, which assumes the  form 
\be 
\ddot{X}(t) + 4\dot{X}(t) + 3X(t) + a\frac{X(t)}{1 + X^2(t)}=0.
\label{eq4} 
\ee 

\par  This representation for the gap equation in the walking regime was first obtained by Cohen and Georgi in Ref.\cite{georgi} and corresponds to the  equation of a unit  mass subjected to the anharmonic potential   
$$V(X) = \frac{1}{2}\left[3X^2(t) + a\ln\left( 1 + X^2(t) \right)\right],$$
which is  quadratic  with  a  logarithmic correction  due  to the $SU(N)$ gauge theory. In the limit of small and large  $X(t)$  the  potential  is  approximately  harmonic, and in these limits the criticallity condition of Eq.(\ref{eq4}) can be analyzed, making analogy with the critical behavior shown by a damped harmonic oscillator subjected to the boundary conditions in the infrared (IR)[$t=t_0$] and ultraviolet (UV)[$t=t_{\Lambda}$]  
regions\cite{Richard,georgi} 
\br 
&& \lim_{t\rightarrow t_0} \frac{\dot{X}(t)}{X(t)} = -1 \nonumber \\
&& \lim_{t\rightarrow t_{\Lambda}} \frac{\dot{X}(t)}{X(t)} = -3. 
\label{eq5} 
\er 

\par The solution of the corresponding linearized equation [Eq.(\ref{eq4})] for $ a < 1$ is described by
\be 
 X(t)  = Ae^{-(2 + \sqrt{1-a})t} + Be^{-(2 - \sqrt{1-a})t} = Ae^{-(3-\gamma_m)t} + Be^{-(1+\gamma_m)t} ,     
\ee 
\noindent  where $\gamma_m = 1 - \sqrt{1-\frac{\alpha}{\alpha_c}}$ is the  mass anomalous dimension of the quark condensate $\langle \bar{Q} Q\rangle$, $\alpha = \frac{g^2}{4\pi}$, $\alpha_c = \frac{\pi}{3C_2}$  and $a = \frac{\alpha}{\alpha_c}$. 

\par Eq.(\ref{eq4}) is described by a damped harmonic oscillator  in the limit of small
$X(t)$, which corresponds to the known behavior of the gap equation solution in the asymptotic region, $t \to t_{\Lambda}$, obtained 
for $a <1$. According to Ref.\cite{georgi} precisely in this case OPE  provides an interpretation  of  the  parameters  appearing  in  the asymptotic  solution of the  gap  equation. Moreover, dynamical chiral symmetry breaking 
does not occur for $a <1$, on the other hand it is possible to investigate the critical behavior of this gap equation when $a \to 1$ with the following  transformation \cite{georgi}
\be 
Y(t) = e^{(1+\gamma_m)t}X(t)
\label{eq6} 
\ee
 
\par With the new coordinate shown in Eq.(\ref{eq6}) we can verify the following relation between  $X(t)$ and $Y(t)$
\be 
\frac{\dot{X}(t)}{X(t)} = -(1+\gamma_m) + \frac{\dot{Y}(t)}{Y(t)}. 
\label{eq7} 
\ee 
Now the differential equation satisfied by $Y(t)$  takes the form\cite{georgi}
\be
\ddot{Y}(t) + 2\sqrt{1-a}\dot{Y}(t) -a\frac{Y^3(t)}{Y^2(t) + e^{2(1+\gamma_m)t}}=0
\label{eq8} 
\ee 
\noindent  and  the boundary conditions for  $Y(t)$ in  the infrared (IR)[$t=t_o$] and ultraviolet (UV)[$t=t_{\Lambda}$]  regions can be obtained from Eq.(\ref{eq7}), leading to
\br 
&& \lim_{t\rightarrow t_0} \frac{\dot{Y}(t)}{Y(t)} = \gamma_m \nonumber \\
&& \lim_{t\rightarrow t_{\Lambda}} \frac{\dot{Y}(t)}{Y(t)} = \gamma_m -2. 
\label{eq9} 
\er 

\par The boundary conditions in the $Y(t)$ coordinate  reflect  the  expected behavior of the dynamical fermion  mass generated
 by the condensate $\langle \bar{Q} Q\rangle$ in the infrared (IR)[$t=t_o$] and ultraviolet (UV)[$t=t_{\Lambda}$] limits. As
observed by Cohen and Georgi: ``chiral symmetry breaking resides not in the solutions to the gap equation, but in
the boundary conditions" \cite{georgi}. For example, if we include the  asymptotic  freedom behavior into the gap equation, 
i.e. the running charge $(a \to a(t))$, in the (UV) limit  $(a(t_{\Lambda}) \to 0)$  we have  $\gamma_m \to 0$, and
$\frac{\dot{Y}(t_{\Lambda})}{Y(t_{\Lambda})} \approx -2$ leads to  
$$
m_f  \sim \mu^3\Lambda^{-2}. 
$$
In the case of a large number of fermions (a walking theory) , where $\gamma_m \approx 1$, we have 
$\frac{\dot{Y}(t_{\Lambda})}{Y(t_{\Lambda})} \approx -1$ and in this case 
$$
m_f \sim \mu^2\Lambda^{-1}. 
$$

\par These examples illustrate as  the boundary conditions, in the $Y(t)$ coordinate, are affected by the inclusion of effects like a large number of fermions (walking),  or any interaction that  modifies the behavior of the fermionic condensate $\langle \bar{Q} Q\rangle$. 
Looking at Eq.(\ref{eq9}) it is still an open problem to find a viable phenomenological model where we may have a large $\gamma_m$ in
the perturbative Banks and Zaks scenario \cite{BZ}. As we shall discuss in the following, the inclusion
of four-fermion interactions will modify the boundary conditions presented in Eq.(\ref{eq9}) in the (UV) region, and the new conditions will show clearly the possibility of a large range of $\gamma_m$ values depending on the behavior of the new coupling constants.
 
\par Eq.(\ref{eq3}) can be also represented by 
\be
X(t)=  \frac{a}{2}e^{-3t}I_1(t) + \frac{a}{2}e^{-t}I_2(t)
\label{eq10} 
\ee 
\noindent where we identify 
\br 
&& I_1(t) = \int^{t}_{t_0}dse^{3s}\frac{X(s)}{1+X^2(s)} \nonumber \\
&& I_2(t) =\int^{t_{\Lambda}}_{t}dse^{s}\frac{X(s)}{1+X^2(s)} \nonumber 
\er 
\noindent  and if we include a four-fermion interaction we have a new contribution to the gap equation 
\br 
X_{4f}(t_{\Lambda})=  ge^{-3t_{\Lambda}}\int^{t_{\Lambda}}_{t_0}dse^{3s}\frac{X(s)}{1+X^2(s)} = ge^{-3t_{\Lambda}}I_1(t_{\Lambda})
\er 
\noindent with $g = \frac{G\Lambda^2}{4\pi^2}$. The incorporation of the  four-fermion interaction produces the following change in the boundary condition for  $X(t)$ in  the ultraviolet region 
\be 
X(t_{\Lambda}) = -\frac{1}{3}\dot{X}(t_{\Lambda}) - \frac{2g}{3a}\dot{X}(t_{\Lambda}) .
\ee 

\par Considering Eq.(\ref{eq7}) the new boundary condition for the $Y(t)$ coordinate in the ultraviolet region becomes
\br 
&& \left(\frac{\dot{Y}(t_{\Lambda})}{Y(t_{\Lambda})}\right)_{4F} = \frac{(\gamma_m -2) + \frac{2g}{a}(1+\gamma_m)}{1+\frac{2g}{a}}. 
\label{eq11} 
\er 
\noindent The four-fermion interaction becomes relevant in the (UV) region when  $t = t_{\Lambda}$ and  the condition that determines the critical line for the  gap equation [Eq.(\ref{eq8})], when $a\to 1$, corresponds to  $\ddot{Y}(t_{\Lambda}) = 0$ in such a way that the boundary condition given by Eq.(\ref{eq11}) leads to
\br 
-(1+\omega) + 2g(2 - \omega) = 0
\label{eq12}
\er
where $\omega =  \sqrt{1-\frac{\alpha}{\alpha_c}}$. Therefore, from Eq.(\ref{eq12}), when $a\to 1$ we can determine the critical line which separates the symmetric and asymmetric chiral phases of a model with relevant four-fermion interaction; and this critical line is given by 
\be 
 g = \frac{(1+\omega)}{2(2 - \omega)}  , 
\label{eq13}
\ee 
\par Usually the critical line is determined from the asymptotic expansion of the hypergeometric function associated to the solution of the gap equation for $\Sigma(p)$, Eq.(\ref{eq1}),  after replacing this result into the ultraviolet boundary condition,  
giving the well known result \cite{yama1, yama2, mira2, yama3, mira3, yama4}
\be 
g = \frac{1}{4}(1+\omega)^2 .
\label{eq14}
\ee 
\noindent For the purpose of comparison we show in Fig.(1) the behavior of the critical line obtained from the above equations 
\begin{figure}[!ht]
\begin{center}
\hspace*{2cm}\epsfig{file=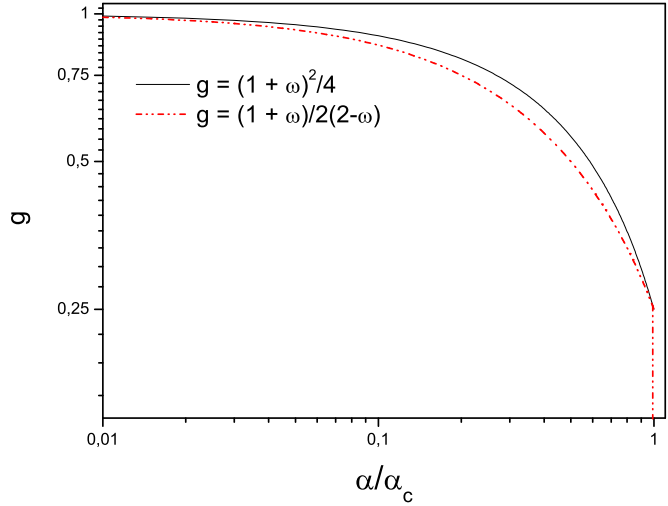,width=1\textwidth}
\vspace*{-4cm}
\caption{ Behavior of the critical line obtained from  Eqs. (\ref{eq13}) and (\ref{eq14})   as a function of $\frac{\alpha}{\alpha_c}$,
where $\omega=\sqrt{1-\frac{\alpha}{\alpha_c}}$.}
\end{center}
\end{figure}

\par We emphasize that in our case we do not use the knowledge about the asymptotic expression assumed by $\Sigma(p)$,  as usually
performed to obtain Eq. (\ref{eq14}). The determination of the critical line in our approach is only due to the modifications in the form assumed by the boundary condition $\dot{Y}(t)/Y(t)$ as $t \to t_{\Lambda}$, due to the  presence of an additional four-fermion
interaction and the fact that the criticallity condition in this case is given by $\lim_{t\rightarrow t_{\Lambda}} \ddot{Y}(t)=0$.
\noindent Therefore in Figure 1 the two critical lines do not exhibit exactly the same behavior, on the other hand, in the extremes
delimited  by the (IR) and (UV) conditions this line accurately reflect the behavior of how the  mass anomalous dimension of the fermionic condensate is modified by the inclusion of new interactions. 

\par  As we mentioned earlier, our intention was to verify how the mass anomalous dimensions may vary with the different boundary conditions after the inclusion of effects like the four-fermion interactions, therefore assuming the result described by Eq.(\ref{eq11}), it is possible
verify that for $g \approx 1$  
\be 
\gamma_m(t_{\Lambda}) =  \left(\frac{\dot{Y}(t_{\Lambda})}{Y(t_{\Lambda})}\right)_{4F} \approx  2 + \frac{\dot{Y}(t_{\Lambda})}{Y(t_{\Lambda})}
\label{eq15}
\ee
\noindent so that $\frac{\dot{Y}(t_{\Lambda})}{Y(t_{\Lambda})}  =  0$ when  the four-fermion interactions becomes relevant and in this case 
$\gamma_m = 2$. Thus, in order to verify how $(\gamma_m)$ is modified by changes in the boundary conditions, we will consider the dynamical behavior of  
\be 
\gamma_m(t) = 2 + \frac{\dot{Y}(t)}{Y(t)}
\label{eq16}
\ee 
\noindent and compute the numerical solution of the fermionic  gap equation  in the  anharmonic oscillator representation, Eq.(\ref{eq8}) for $a \to 1$ , with the different set of UV boundary conditions indicated by $C_i$, where 
\be 
C_i = \left(\frac{\dot{Y}(t_{\Lambda})}{Y(t_{\Lambda})}\right)_i
\ee 
\noindent  and $i=$ stands for OPE, walking (Walk) and four-fermion interaction (4F), with the limits $C_{\rm{OPE}} = -2$, $C_{\rm{Walk}} = -1$ and $C_{\rm{4F}} = 0$ (see Eqs.(\ref{eq9}) and (\ref{eq15})). In the  Fig.(2) we show the behavior of  Eq.(\ref{eq16}) for this set of different (UV) boundary conditions 
\begin{figure}[!ht]
\begin{center}
\hspace*{2.3cm}\epsfig{file=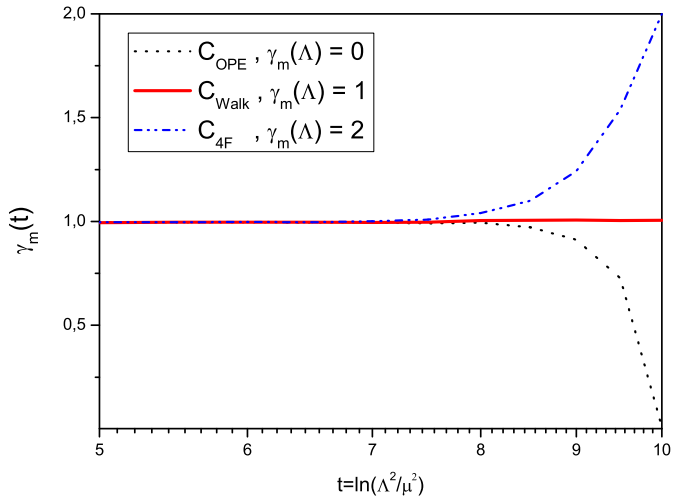,width=1\textwidth}
\vspace*{-4cm}
\caption{ Behavior of Eq.(\ref{eq16}) ($\gamma_m (t)$) for the set of different (UV) boundary conditions $C_i$, $C_{\rm{OPE}} = -2$, $C_{\rm{Walk}} = -1$ and $C_{\rm{4F}} = 0$.}
\end{center}
\end{figure}

\par As mentioned at the beginning of this note there are many determinations of the mass anomalous dimension of $SU(N)$ models. Considering
the different $\gamma_m$ values obtained in the literature, despite the different methods to obtain this quantity, it is interesting to have a simple explanation of the origin of these values. 
In this note we discussed  how the boundary conditions of SDE in the  anharmonic oscillator representation are directly related  with  the mass anomalous dimensions,  and how such conditions are  affected by inclusion of effects like a large number of fermions  or by inclusion of four-fermion interactions. 

\par The fact that the introduction of four-fermion interactions induce large anomalous dimensions is known in the literature for a long time in what is now known as gauged Nambu-Jona-Lasinio  models, see  Refs.\cite{yama1, yama2, mira2, yama3, mira3, yama4}. Essentially we just recovered known results in a different approach. Moreover, at this point we should mention that one of the contributions of this work, compared with the previous  studies, is that  we have shown how the mass anomalous dimensions are just dictated by the boundary conditions. 

\par In Fig. (2) we illustrate how the effect of different UV boundary conditions produces a distinct behavior for $(\gamma_m)$. This  is a simple result, although we are not aware that this fact was stated before. We also recover the behavior of the critical line obtained in the context of these models, however this result is obtained without using the knowledge about the asymptotic $\Sigma(p)$ expressions,  as usually performed to obtain Fig(1). The determination of the critical line in our approach is only due to the modifications in the form assumed by the boundary conditions  due to the  presence of an additional four-fermion
interaction.

\section*{Acknowledgments}
This research was  partially supported by the Conselho Nacional de Desenvolvimento Cient\'{\i}fico e Tecnol\'ogico (CNPq) and by grant 
2013/22079-8 of Funda\c{c}\~{a}o de Amparo \`{a} Pesquisa do Estado de S\~ao Paulo (FAPESP).





\end{document}